\documentclass[USenglish,oneside,twocolumn]{article}
\usepackage[big]{dgruyter_NEW}

\usepackage{listings}          % code
\usepackage[caption=false]{subfig} %only subfigure environment accepted by the establishment
\usepackage{amsfonts,amssymb} %math
\usepackage{times}
\usepackage{enumerate}
\usepackage{marginnote}
\usepackage{color}
\usepackage{url}
\usepackage{algorithm}% http://ctan.org/pkg/algorithms
\usepackage{algpseudocode}
\usepackage{setspace}
\newcommand{\remove}[1]{{}}  % removes a portion of text

\newcommand{\calx}{\mathcal{X}}

\newcommand{\calz}{\mathcal{Z}}

\newcommand{\calp}{\mathcal{P}}

\newcommand{\calg}{\mathcal{G}}

\newcommand{\K}{K}

\newcommand{\ind}{\mathrm{cover}}
\newcommand{\rcountry}{{r_\mathrm{large}}}
\newcommand{\rcity}{{r_\mathrm{small}}}

\newcommand{\dx}{{d_{\scriptscriptstyle\calx}}}

\newcommand{\euclid}{{d_{\mathrm{euc}}}}

\newcommand{\binary}{{d_{\mathrm{bin}}}}
\newcommand{\threshold}{{d_r}}

\newcommand{\hamming}{{d_h}}

\newcommand{\dprob}{{d_\calp}}

\newcommand{\reals}{\mathbb{R}}

\newcommand{\priv}[1]{$#1$-privacy}

\newcommand{\Geoind}{Geo-in\-dis\-tin\-gui\-sha\-bi\-li\-ty}
\newcommand{\geoind}{geo-in\-dis\-tin\-gui\-sha\-bi\-li\-ty}
\newcommand{\egeoind}[1][\epsilon]{$#1$-geo-in\-dis\-tin\-guisha\-bil\-ity}

\renewcommand{\div}{\ \mathrm{div}\ }

\newcommand{\smallsum}[1]{\textstyle{\sum_{#1}\:}}

\newcommand{\smallfrac}[2]{\textstyle{\frac{#1}{#2}}}

\newcommand{\req}{\mathrm{req}}          %requirement
\newcommand{\pp}{m}          % privacy mass
\newcommand{\pl}{PL}
\newcommand{\mathpl}{\mathrm{PL}}
\newcommand{\mathemm}{\mathrm{EM}}

\newcommand{\emm}{EM}
\DeclareMathOperator*{\argmin}{arg\,min}

\newcommand{\da}{d_A}

\newcommand{\adv}{\textsc{AdvError}}
\newcommand{\advarg}[4]{\adv(#1, #2, #3, #4)}

\title{Constructing elastic distinguishability metrics for location privacy}
\author[1]{Konstantinos Chatzikokolakis}
\affil[1]{CNRS and LIX, \'{E}cole Polytechnique, France}
\author[2]{Catuscia Palamidessi}
\affil[2]{INRIA and LIX, \'{E}cole Polytechnique, France}
\author[3]{Marco Stronati}
\affil[3]{LIX, \'{E}cole Polytechnique, France}

\allowdisplaybreaks[1]

\begin{document}

\maketitle
\begin{abstract}
WWith the increasing popularity of hand-held devices, location-based
applications and services have access to accurate and real-time
location information, raising serious privacy concerns for their
users.
The recently introduced notion of \emph{\geoind{}}
tries to address this problem by adapting the well-known concept of
differential privacy to the area of location-based systems.
% This is actually an instance of a more general property called
% \priv{\dx}, defined on arbitrary metric spaces:
% the notion of \geoind{} is obtained by considering the
% Euclidean metric.

Although \geoind{} presents various appealing aspects,
it has the problem of treating space in a uniform way, imposing the
addition of the same amount of noise everywhere on the map.
%This results in a non-optimal trade-off between privacy and utility,
%because low-density areas require more obfuscation than high-density
%areas to achieve the same degree of privacy.
In this paper we propose a novel elastic distinguishability metric that warps the
geometrical distance, capturing the different degrees of density of each
area. As a consequence, the obtained mechanism
adapts the level of noise while achieving the same degree
of privacy everywhere.
% The elastic metric can be efficiently computed and, in combination with
% the \priv{\dx} framework, achieves the desired degree of privacy
% everywhere with the minimal amount of noise locally required for
% ``hiding-in-the-crowd''.
% Another problem (common to all location privacy approaches based on
% random noise) is that the repeated application of the mechanism will
% eventually disclose the frequently exposed locations, such as work or
% home.
We also show how such an elastic metric can easily incorporate the concept
of a ``geographic fence'' that is commonly employed to protect
the highly recurrent locations of a user, such as his home or work.

We perform an extensive evaluation of our technique by building an elastic
metric for Paris' wide metropolitan area, using semantic information from the
OpenStreetMap database. We compare the resulting mechanism against the Planar
Laplace mechanism satisfying standard \geoind, using two real-world datasets
from the Gowalla and Brightkite location-based social networks. The results show
that the elastic mechanism adapts well to the semantics of each area, adjusting
the noise as we move outside the city center, hence offering better overall
privacy.
\footnote{
  This work was partially supported by the European Union 7th FP project MEALS, by the project ANR-12-IS02-001 PACE, and by the INRIA Large Scale Initiative CAPPRIS.}
\end{abstract}

\journalname{Proceedings on Privacy Enhancing Technologies}
\DOI{10.1515/popets-2015-0023}
\startpage{1}
\received{2015-02-15}
\revised{2015-05-13}
\accepted{2015-05-15}

\section{Introduction}

The availability of devices capable detecting the geographical
position with pretty good accuracy (e.g. wifi-hotspots, GPS, etc.)
has led to a proliferation of applications that use the location data
to provide a range of services.
These applications, called location-based services (LBSs), include
points-of-interest retrieval, coupon providers, GPS navigation,
location aware social networks, etc. 
While the value of these services is undeniable, as attested by their
vast popularity, at the same time there are growing concerns about the
potential privacy breaches that the user is exposed to, due to the
constant disclosure of location information.

Among the various approaches proposed in the literature to address the
problem of privacy in the use of LBSs, those based on obfuscating the
real location by adding random
noise~\cite{Shokri:11:SP,Shokri:12:CCS,Andres:13:CCS,Fawaz:14:CCS}
have emerged as the most robust to side-information-enhanced attacks.
Additionally, the \emph{\geoind{}}
framework~\cite{Andres:13:CCS,Fawaz:14:CCS} has the appealing features
of providing formal privacy guarantees independent from the user's
prior, being robust with respect to combination attacks, and offering
a good trade off between privacy and utility.
These features are inherited from the framework of differential
privacy~\cite{Dwork:06:ICALP} which prevents an adversary from distinguishing
datasets that are ``close'' based on the Hamming metric.
\Geoind{} follows the same principles while
using the Euclidean distance between locations.
%; similarly we could
%use any distinguishability metric $\dx$ that\priv{\dx}
%\cite{Chatzikokolakis:13:PETS}, \priv{\dx}, where
%the distance is the Euclidean distance.

Formally, a mechanism provides \geoind{} if the
user's real location $x$ is \emph{indistinguishable} from other nearby
locations $x'$, meaning that the mechanism should report any noisy
location $z$ with a probability similar to that of reporting the same
$z$ when the real location is $x'$.
The level of similarity depends on the geographical distance between
$x$ and $x'$.
More precisely, the ratio between the two probabilities is bound by an
expression that grows exponentially with the distance.
This means that the mechanism protects the accuracy of the real
location but reveals that the user is, say, in Paris instead than
London, which is appropriate for the kind of applications (LBSs) we
are targeting.
\Geoind{} is typically achieved by reporting a location
obtained from the user's location by adding noise drawn from a
2-dimensional Laplace distribution.

It is to be noted that the intuition behind most notions of privacy is
that of \emph{hiding in a crowd}.
The ``crowd'' may be of different nature: we might want to hide among
several people, hide the kind of shop we visit,
the activity we perform, etc.
\Geoind{} provides this property in the context of the
information that can be inferred from the location. By properly
setting a parameter of the definition, one can establish the area of
the points indistinguishable from the real location, that is, how many
elements (resident people, shops, recreational centers etc.) are made
indistinguishable from the ones in the real location.

However, one problem with the \geoind{} framework is that, being based on the
Euclidean distance, its protection is uniform in space, while the density of the
elements that constitute the ``crowd'' in general is not. This means that, once
the privacy parameter is fixed, a mechanism providing \geoind{} will generate
the same amount of noise independently of the real location on the map, i.e.,
the same protection is applied in a dense city and in a sparse countryside. As a
consequence, an unfortunate decision needs to be made: one could either tune the
mechanism to the amount of noise needed in a dense urban environment, leaving
less dense areas unprotected. Or, to ensure the desired level of privacy in
low-density areas, we can tune the mechanism to produce a large amount of
noise, which will result in an unnecessary degradation of utility in
high-density areas.

The idea we pursue in this paper is to adopt a privacy notion that reflects the
local characteristics of each area. This can be achieved while maintaining the
main principles of \geoind{}, by replacing the Euclidean distance
with a constructed \emph{distinguishability metric} $\dx$. The resulting notion,
called \priv{\dx} in \cite{Chatzikokolakis:13:PETS}, ensures that secrets which
are close wrt $\dx$ should remain indistinguishable, while secrets that are
distant wrt $\dx$ are allowed to be distinguished. We then have the flexibility
to adapt the distinguishability metric $\dx$ to our privacy needs.

Going back to the intuition of privacy as being surrounded by a crowd, we can
reinterpret it in the light of distinguishability metrics. On one hand people or
points of interest can be abstracted as being a \emph{privacy mass} that we can
assign differently to every location. On the other hand the concept of being
close to a location rich in privacy can be seen as the desire to be similar, or
indistinguishable to such a location. Therefore we can express our intuitive
privacy with a distinguishability metric that satisfies the following
requirement: every location should have in proximity a certain amount of privacy
mass. This can be better formalized with a requirement function $\req(l)$ that for every
distinguishability level $l$, assigns a certain amount of privacy mass that must
be present within a radius $l$ in the metric $\dx$.
Contrary to \geoind{}, that considers space uniform and assigns to every
location the same privacy value, we build a metric that is flexible and adapts
to a territory where each location has a different privacy importance. In
comparison with the Euclidean metric, our metric stretches very private areas
and compresses the privacy poor ones in order to satisfy the same requirement
everywhere, for this reason we call it an \emph{elastic metric}.
By using \priv{\dx} with a metric that takes into account the
semantics of each location, we preserve the strengths of the
geo-indistinguishability framework while adding flexibility,
borrowing ideas from the line of work of $l$-diversity
\cite{Xue:09:LoCa,Machanavajjhala:07:TKDD}.
This flexible
behavior reflects also on the utility of the resulting mechanism, areas poor in
privacy will result in more noisy sanitization.

We then need a way to compute the actual metric $\dx$ satisfying the requirement
$\req(l)$. We propose a graph-based algorithm that can efficiently compute an
elastic metric for a large number of locations. The algorithm requires a set of
locations marked with privacy mass and a privacy requirement to satisfy.
Starting from an empty graph, it iteratively adds weighted edges to satisfy the
requirement. The resulting distance between two locations is the weight of the
shortest path connecting them. Once obtained our metric we show how to use an
exponential distribution to obtain automatically a $\dx$-private mechanism that
can also be efficiently implemented.

Another problem that arises in the approaches to location privacy
based on random noise -- and \geoind{} is not immune to
it -- is that the reiterate use of the mechanism will eventually
disclose the frequently visited locations, such as home or office.
% This is due to the law of the large numbers: the disposition of the
% reported locations will eventually show peaks on the points (real
% locations) in which the user activates the mechanism with high
% frequency.
A solution (commonly employed for various privacy purposes) consist in building
a ``fence'' around sensitive locations so that all points inside are completely
indistinguishable from each other. In this way the attacker will be able, after
many iterations, to identify the fence but not the exact location inside the
fence.
%In a sense, instead of having our sensitive locations exactly
%determined during the use of the mechanism, we rather declare publicly
%beforehand their approximate position.
We show that such a solution can be elegantly expressed in the
distinguishability metric $\dx$, and can be easily incorporated in our
algorithm.
%compositional with other metrics we may decide to use outside the
%fence.
%The fenced metric effectively stops the linear growth of the privacy
%budget and allows for a prolonged use of the mechanism.

Finally, we show the applicability of our technique by evaluating on two
real-world datasets. We start by building an elastic metric for Paris' wide
metropolitan area, in a grid of $562,500$ locations covering an area of $5600$
km$^2$. Privacy mass is computed from semantic information extracted from the
OpenStreetMap database, and the whole computation is performed in under a day
with modest computational capability, demonstrating the scalability of the
proposed algorithm.

We then compare the elastic mechanism to the Planar Laplace mechanism satisfying
\geoind{}, on two large areas in the center of Paris as well as a
nearby suburb. The evaluation is performed using the datasets of Gowalla and
Brightkite\cite{Cho:11:SIGKKD}, two popular location-based social networks, and
the widely used Bayesian privacy and utility metrics of Shokri et al.
\cite{Shokri:11:SP}. The results show that the dynamic behavior of the elastic
mechanism, in contrast to Planar Laplace, provides adequate privacy in both high
and low-density areas.

\subsubsection*{Contributions}
\begin{itemize}
	\item We propose the use of elastic metrics to solve
		the flexibility problem of \geoind{}. We formalize a requirement of such
		metrics in terms of privacy mass, capturing properties such as space,
		population, points of interest, etc.
	\item We propose an efficient and scalable
  		graph-based algorithm to compute a metric $\dx$ satisfying this requirement.
	\item We show that the technique of geo-fences can be elegantly expressed in
		the metric and incorporated in the algorithm.
	\item We perform an extensive evaluation of our technique in a large
		metropolitan area
		using two real-world datasets, showing the advantages of the elastic
		metric compared to standard \geoind{}.
\end{itemize}

\paragraph{Plan of the paper}
In the next section we recall some preliminary notions about
\priv{\dx} and \geoind{}.
In Section \ref{sec:elastic_metric} we present in detail the elastic
metric.
First how to extract meaningful privacy resource for each location,
then the definition of a privacy requirement and finally the
graph-based algorithm that generated the metric.
In Section \ref{sec:fenced_metric} we describe how to model
geographical fences with a metric and how to integrate it with the
elastic metric algorithm.
Finally in Section \ref{sec:evaluation} an elastic mechanism is built
and evaluated in comparison with a geo-indistinguishable mechanism.

%%% Local Variables: 
%%% mode: latex
%%% TeX-master: "paper"
%%% End: 

\section{Preliminaries}\label{sec:preliminaries}

We briefly recall here some useful notions from the literature.

\paragraph*{Probabilistic model}

We first introduce a simple model used in the rest of the paper. We start with a
set $\calx$ of \emph{points of interest} (i.e. a subset of $\reals^2$), typically the user's possible
locations. Moreover, let $\calz$ be a set of possible \emph{reported
locations}, which typically coincides with $\calx$, although this is not
necessary. 
For the needs of this paper we consider $\calx,\calz$ to be finite.

The selection of a reported value $z\in \calz$ is \emph{probabilistic}; $z$ is
typically obtained by adding random noise to the actual location $x$.
The set of probability distributions over $\calz$ is denoted by $\calp(\calz)$.
We define the \emph{multiplicative distance} between two distributions
$\mu_1,\mu_2\in\calp(\calz)$ as
$
	\dprob(\mu_1,\mu_2) \allowbreak= \allowbreak \sup_{Z\subseteq\calz}| \ln \smallfrac{\mu_1(Z)}{\mu_2(Z)}|
$,
with the convention that $|\ln\smallfrac{\mu_1(Z)}{\mu_2(Z)}|=0$ if both
$\mu_1(Z),\mu_2(Z)$ are zero and $\infty$ if only one of them is zero.

A \emph{mechanism} is a (probabilistic) function $K:\calx\to\calp(\calz)$
assigning to each location $x\in\calx$ a probability distribution on $\calz$,
where $\K(x)(z)$ is the probability to report
$z \subseteq \calz$, when the user's location is $x$.

\paragraph*{Geo-indistinguishability and \priv{\dx}}
The notion of \emph{\geoind{}} was proposed in
\cite{Andres:13:CCS} as an extension of differential privacy in the area of
location privacy. The main idea behind this notion is that it forces the output
of the mechanism applied on locations $x,x'$, i.e. the distributions
$K(x),K(x')$, to be similar when $x$ and $x'$ are geographically close,
preventing an adversary from distinguishing them, while it relaxes the
constraint when $x,x'$ are far away from each other, allowing a service provider
to distinguish points in Paris from those in London.
%The complete definition can be written as follows:
Let $\euclid(\cdot,\cdot)$ denote the Euclidean metric; a mechanism $\K$
satisfies \egeoind{} iff for all $x,x'$:
\[
	\dprob(\K(x),\K(x')) \le \epsilon\euclid(x,x')
\]
Equivalently, the definition can be formulated as
$
	\K(x)(z) \le $ $\allowbreak e^{\epsilon \euclid(x,x')} \allowbreak \K(x')(z)
$
for all $x,x'\in\calx,z\in\calz$.

The quantity $\epsilon \euclid(x,x')$ can be viewed as the
\emph{distinguishability level} between the secrets $x$ and $x'$. The use of the
Euclidean metric $\euclid$ is natural for location privacy: the \emph{closer}
(geographically) two points are, the \emph{less distinguishable} we would like them to
be. Note, however, that any other (pseudo-)metric could be used instead of $\euclid$, such
as the Manhattan metric or driving distance, depending on the application. The
definition that we obtain by using an arbitrary distinguishability metric $\dx$,
i.e. requiring that $\dprob(\K(x),\K(x')) \le \dx(x,x')$, is referred to as
\priv{\dx}, and is studied on its own right in \cite{Chatzikokolakis:13:PETS}.
In Section~\ref{sec:elastic_metric} it is argued that a properly constructed $\dx$ can
provide location privacy while taking into account the semantic properties of
each location.

It should be noted that standard differential privacy simply corresponds to
\priv{\epsilon \hamming(x,x')}, where $\hamming$ is the Hamming distance between
databases $x,x'$, i.e. the number of individuals in which they differ.
If $x,x'$ are \emph{adjacent}, i.e. they differ in a single individual, then
$\hamming(x,x')=1$ and the distinguishability level between such databases
is exactly $\epsilon$.

Two characterization results are also given in
\cite{Andres:13:CCS,Chatzikokolakis:13:PETS},
providing intuitive interpretations of \geoind{} and \priv{\dx}.

It should be emphasized that \geoind{} aims at protecting the user's
\emph{location}, not his \emph{identity} (protecting the link between a person
identity and its location is the goal of several techniques in the location
privacy literature, see Section~\ref{sec:related-work}). In our setting a user
might even be authenticated with the LBS, for instance to obtain personalized
search results, while wishing to protect his location.

\paragraph*{Distinguishability level and $\epsilon$}

A point worth emphasizing is the role of the distinguishability level, and its
relationship to $\epsilon$ in each definition. The distinguishability level
between two secrets $x,x'$ is their distance in the \emph{complete} privacy
metric employed, that is $\epsilon\hamming(x,x')$ for differential privacy,
$\epsilon\euclid(x,x')$ for \geoind{}, and $\dx(x,x')$ for
\priv{\dx}. Secrets that are assigned a ``small'' distinguishability level will
remain indistinguishable, providing privacy, while secrets with a large
distance are allowed to be distinguished in order to learn something from the system.
Typical values that are considered ``small'' range from $0.01$ to $\ln 4$;
we denote a small distinguishability level by $l^*$, and in our evaluation
we use $l^*= \ln 2$.

In the case of differential privacy, $\epsilon$ is exactly the
distinguishability level between adjacent databases (since $\hamming(x,x')=1$
for such databases), hence we directly use our ``small'' level $l^*$ for $\epsilon$.
For \geoind{}, however, $\epsilon$ represents the
distinguishability level for points such that
$\euclid(x,x')=1$; however, depending on the unit of measurement as well as on
the application at hand, we might or might not want to distinguish points at
a unit of distance. To choose $\epsilon$ in this case, we start by defining a radius
$r^*$ of \emph{high protection}, say $r^* = 300$ meters for an LBS application
within a big city, and we set $\epsilon = l^*/r^*$. As a consequence,
points within $r^*$ from each other will have distinguishability level at
most $l^*$, hence an adversary will be unable to distinguish them, while points
will become increasingly distinguishable as the geographic distance between them
increases.

Note the difference between $\euclid(x,x')$, the geographic distance between
$x,x'$, and $\epsilon\euclid(x,x')$, the distinguishability level between
$x,x'$. In other words, $\epsilon$ stretches the Euclidean distance turning it
into a distinguishability metric. To avoid confusion, throughout the paper we
use $r$ to denote geographical distances and radii, and $l$ to denote
distinguishability levels.

In the case of \priv{\dx}, $\dx(x,x')$ directly gives the distinguishability
level between $x$ and $x'$. In Section~\ref{sec:elastic_metric} we investigate some
properties that such a metric should satisfy in order to provide location
privacy while taking into account the semantics of each location. Then, in
Section~\ref{algo}, we propose a graph-based algorithm for constructing such
a metric.

\paragraph*{Repeated application}
Any obfuscation mechanism is bound to cause privacy loss when
used repeatedly. In the case of an $\epsilon$-geo-indistinguishable mechanism
$K$, applying it $n$ times will satisfy \egeoind[n\epsilon]. This is typical
in the area of differential privacy, in which $\epsilon$ is thought as a budget
which is consumed with every query.

The situation is similar in the case of \priv{\dx}: applied $n$ times, a
mechanism will satisfy \priv{n\dx}. This means that the distinguishability
level between $x,x'$ after $n$ applications is $n\dx(x,x')$; if $\dx(x,x')>0$
then as $n$ grows $x$ and $x'$ are bound to become completely distinguishable.
However, if we use a pseudo-metric such that $\dx(x,x')=0$, then $x,x'$ are
completely indistinguishable, and will remain so under any number of repetitions
$n$. This property will be exploited by the ``fence'' technique of
Section~\ref{sec:fenced_metric}.

\paragraph*{Planar Laplace and Exponential mechanisms}

The typical approach for achieving differential privacy is adding noise
from some sort of Laplace-like distribution. In the case of the Euclidean
distance, the \emph{Planar Laplace (\pl)} mechanism \cite{Andres:13:CCS} can be
employed. When applied on location $x$, this mechanism draws a location $z$
from the continuous plane with probability density function:
\[
	\frac{\epsilon}{2\pi} e^{- \epsilon\euclid(x,z)}
\]
In \cite{Andres:13:CCS} a method to efficiently draw from this distribution is
given, which uses polar coordinates and involves drawing an angle and a radius
from a uniform and a gamma distribution respectively. The mechanism can be
further discretized and truncated, and can be shown to satisfy \egeoind.

Furthermore, in the case of an arbitrary distinguishability metric $\dx$, a
variant of the
Exponential mechanism \cite{McSherry:07:FOCS} can be employed. When applied at location $x$, this mechanism
reports $z$ with probability:
\[
	c_x e^{-\frac{1}{2} \dx(x,z)}
	\quad\text{with}\quad c_x = (\smallsum{z'} e^{-\frac{1}{2}\dx(x,z')})^{-1}
\]
where $c_x$ is a normalization factor. This mechanism can be shown to satisfy
\priv{\dx}. Note the difference in the exponent between the two mechanisms:
the Exponential mechanism has a factor $\frac{1}{2}$ missing from the Planar Laplace;
in the proof of \priv{\dx}, this factor compensates for the fact that the
normalization factor $c_x$ is different for every $x$, in contrast to the Planar
Laplace while the normalization factor $\frac{\epsilon}{2\pi}$ is independent
from $x$.
The advantage of this technique is the possibility of obtaining a
privacy mechanism independently of the metric used, allowing us to
focus solely on the metric design.

\paragraph*{Utility}
The goal of a privacy mechanism is not to hide completely the secret but to
disclose enough information to be useful for some service while hiding the rest
to protect the user's privacy. Typically these two requirements go in opposite
directions: a stronger privacy level requires more noise which results in a
lower utility.

Utility is a notion very dependent on the application we target. 
An example of noise insensitive applications are weather forecast
services, where utility remains unchanged even if the reported
location is kilometers away from the real one.
On the contrary an POI research application can tolerate lower noise
addition in order to report meaningful results.

However, to evaluate and compare \emph{general-purpose} location obfuscation
mechanisms, the general principle is to report locations as close as possible
to the original ones. A natural and widely used choice
\cite{Shokri:11:SP,Shokri:12:CCS,Bordenabe:14:CCS} is to define utility as the
\emph{expected geographical distance} between the actual and the reported
locations. Hence, we define the average error of mechanism $K$ on location $x$
as:
\[
	E_K(x) = \displaystyle
		\smallsum{z} K(x)(z) \; \euclid(x,z)
\]
In the case of the Planar Laplace mechanism, the error is independent from $x$
(due to the symmetry of the continuous plane)
and is given by $E_{\mathpl}(x) = 2/\epsilon$.

\section{An elastic distinguishability metric}
\label{sec:elastic_metric}

Rarely we are interested in hiding our geographical location per se,
more commonly we consider our location sensitive because of the many
personal details it can indirectly reveal about us.
For this reason reducing the \emph{accuracy}, through perturbation or
cloaking, is considered an effective technique for reporting a
location that is at the same time meaningful and decoupled from its
sensitive semantic value.
In the preliminaries we explained how in \geoind{} the
privacy level is configured for a specific radius $r^{*}$, that is
perceived as private.
Using $r^{*}=300$ m for a large urban environment is based on the fact
that a large number of shops, services and people can be found within
that radius, limiting the power of inference of the attacker.
This is an intuitive notion of location privacy that we call
\emph{hiding in a crowd}, where the crowd represents the richness and
variety that a location provides to the user's privacy.

As explained in the introduction, the use of $\epsilon\euclid$ as the
distinguishability metric has a major drawback.
The simple use of geographical distance to define privacy ignores the
nature of the area in which distances are measured.
In a big city, $\epsilon$ can be tuned so that strong privacy is
provided within 300 meters from each location but in a rural
environment, they are not perceived as sufficient privacy.
And even inside a city, such a protection is not always adequate:
within a big hospital an accuracy of 300 meters might be enough to
infer that a user is visiting the hospital.

In this section we address this issue using a custom
distinguishability metric that is adapted to the properties of each
area, an \emph{elastic metric}.
More specifically we discuss properties that such a metric should
satisfy, and in the next section we present an algorithm for
efficiently computing such a metric.

Once obtained the elastic metric, we can plug it in the $\dx$-privacy
definition and obtain an \emph{elastic privacy definition} for
location privacy, much like was done for \geoind{}.
Furthermore we can use the Exponential mechanism presented in
Section~\ref{sec:preliminaries} to obtain a \emph{sanitization mechanism}
that satisfies our elastic privacy definition.

\subsection{Privacy mass}\label{sec:mass}

The main idea to overcome the rigidity of \geoind{} is
to construct a distinguishability metric $\dx$ that adapts
depending on the properties of each area.
% In a city, distances can be larger, protecting the user within a
% smaller area, which is sufficient for privacy, while allowing better
% utility.
% In the country, however, distances should become smaller: it might be
% necessary for points within 3 km to be considered indistinguishable to
% achieve reasonable privacy.
However in order to distinguish a city from its countryside, or on a finer
scale, a crowded market place from a hospital, we first need to assign
to each location how much it contributes to the privacy of the user.
In other words we consider \emph{privacy as a resource} scattered on
the geographical space and each locations is characterized by a
certain amount of privacy.

More precisely the privacy of \emph{location $x$} depends on the \emph{points
that are indistinguishable from $x$}. Let $\ind(x)$ denote
the set of points that are ``highly'' indistinguishable from $x$.
For the moment we keep $\ind(x)$ informal, it is properly defined in the next
section. Intuitively, the privacy of $x$ depends:
\begin{enumerate}
	\item on the \emph{number} of points in $\ind(x)$:
		an empty set clearly means that $x$ can be inferred,
		while a set $\ind(x)$ containing a whole city provides high privacy.
		This corresponds to the idea that hiding within a large area provides
		privacy.
	\item on the semantic \emph{quality} of points in $\ind(x)$:
		it is preferable for $\ind(x)$ to contain a variety of POIs and highly
		populated locations, than points in a desert or points all belonging to
		a hospital.
		This corresponds to the idea that hiding within a populated area with a
		variety of POIs provides privacy.
\end{enumerate}

To capture this intuition in a flexible way we introduce the concept of
\emph{privacy mass}. The privacy mass of a location $x$, denoted by $\pp(x)$, is
a number between $0$ and $1$, capturing the location's value at providing
privacy. We also denote by $\pp(A)=\sum_{x \in A}\pp(x)$ the total mass of a set
$A$. The function $\pp(\cdot)$ should be defined in a way such that a set of
points containing a \emph{unit of mass} provides sufficient cover for the user.
Hence, the metric we construct needs to satisfy that
\[
	\pp(\ind(x)) \ge 1 	\qquad \forall x\in\calx
\]

Following the idea that privacy comes by hiding within either a
``large'' or ``rich'' area, we define $\pp(x)$ as
\begin{equation}\label{eq:resource}
  \pp(x) = a + q(x) b
\end{equation}
where $a$ is a quantity assigned to each location simply for ``occupying
space'', $q(x)$ is the ``quality'' of $x$
and $b$ is a normalization factor. 
The quality $q(x)$ can be measured in many
ways; we measure it by querying the OpenStreetMap database for a variety of POIs
around $x$, as explained in Section~\ref{queries}.
Assuming $q(x)$ to be given, we can compute $a$ and $b$ as follows:
we start with the intuition that even in empty space, a user feels
private if he is indistinguishable within some large radius
$\rcountry$, for instance 3000 m ($\rcountry$ can be provided
by the user himself).
Let $B_r(x) = \{ x' \;|\; \euclid(x,x') \leq r \}$ denote the Euclidean ball
of radius $r$ centered at $x$.
Letting $x$ be a location in empty space, i.e. with $q(x)=0$,
intuitively we want that $\ind(x) = B_\rcountry(x)$,
and $\pp(\ind(x)) = 1$, hence
\[
	a = \frac{1}{|B_\rcountry(x)|}
\]
Similarly, in an ``average'' location in a more private place, like a city, a user feels private if he 
is indistinguishable within some smaller radius
$\rcity$, for instance 300 m ($\rcity$ can be also provided
by the user himself). Let
\[
	avg_q = E_x q(B_\rcity(x))
\]
be the
average quality of a $\rcity$ ball (where expectation is taken over
all location in the city). On average we establish that such a ball contains one unit of privacy mass, thus we get:
\begin{align*}
	1 &= a\cdot|B_\rcity(x)|  + b\cdot avg_q
		& \text{hence}
	\\
	b & =\frac{1}{avg_q} (1- \frac{|B_\rcity(x)|}{|B_\rcountry(x)|})
\end{align*}

Note that the intuitive requirement of being indistinguishable from a
set of entities with some semantic characteristics is widely used in the privacy
literature. Most notably, $k$-anonymity
\cite{Kido:05:ICDE,Shankar:09:UbiComp,Bamba:08:WWW,Duckham:05:Pervasive}
requires to be indistinguishable from group of at least $k$ individuals, while
$l$-diversity \cite{Xue:09:LoCa,Machanavajjhala:07:TKDD} adds semantic diversity
requirements: hiding among $k$ hospitals is not acceptable since it still reveals
that we are in a hospital. It should be emphasized, however, that although
we follow this general intuition, we do so inside the geo-indistinguishability
framework, leading to a privacy definition and an obfuscation mechanism
vastly different from those of the aforementioned works,
as explained in more details in Section~\ref{sec:related-work}.

\subsection{Requirement}
Having fixed the function $\pp(x)$, we turn our attention to the
requirement that our distinguishability metric $\dx$ should satisfy in order
to provide adequate privacy for all locations.

Let $B_l(x)$ denote the $\dx$-ball of distinguishability level $l$. The \priv{\dx} property
ensures that, the smaller $l$ is, the harder it will be to distinguish $x$ from
any point in $B_l(x)$. 
Our requirement is that $B_l(x)$ should collect an appropriate amount of privacy
mass:
\begin{equation}\label{eq:requirement}
  \pp(B_l(x)) \geq \req(l) \qquad \forall l \ge 0, x\in \calx
\end{equation}
where $\req(l)$ is a function expressing the required privacy mass at each
level. The algorithm of Section~\ref{algo} ensures that the above property is
satisfied by $\dx$.

It remains to define the $\req(l)$ function. Let $l^*$ denote a ``small'' distinguishability
level (see Section~\ref{sec:preliminaries} for a discussion on
distinguishability levels and what small means. In this paper we use $l^*=\ln
2$). Points in $B_{l^*}(x)$ will be ``highly'' indistinguishable from $x$, hence
$B_{l^*}(x)$ plays the role of $\ind(x)$ used informally in the
previous Section. Privacy mass was defined so that $m(\ind(x)) \ge 1$,
hence we want $\req(l^*) = 1$.

Moreover, as the $\dx$-distance $l$ from $x$ increases, we should
collect even more mass, with the amount increasing quadratically with $l$
(since the number of points increases quadratically).
Hence we define $\req(l)$ as a quadratic function with $\req(0)=0$ and
$\req(l^*) = 1$, that is:
\[
	\req(l) = \big(\frac{l}{l^*}\big)^2
\]

Defining the requirement in terms of privacy mass is a flexible way to adapt it
to the properties we are interested in. Indeed we can re-obtain \geoind{} as a
special case of our new framework if all locations are considered just for their
contribute in space, not quality, i.e. $q(x)=0$. The requirement is then to be
indistinguishable in certain area and in an Euclidean metric it is simply the
area of the circle with radius $l$, a function that is indeed quadratic in $l$.

%In order to determine the steepness of the curve, that is a factor
%inherently dependent on the user expectation of privacy, we let the
%user define a a safe distance $r^{*}$ and we impose to have $\pp^{*}$
%privacy points at that distance.
%In this way we obtain $\req(r) = \frac{\pp^{*}}{{r^{*}}^2} r^2$.
%In our experiments we set $r^{*}$ to $ln 4$ and $\pp^{*}=1$.

\subsection{Extracting location quality}\label{queries}
Our definition of privacy mass depends on the semantic quality
$q(x)$ of a point $x$.
To compute the quality in a meaningful way, we
used the OpenStreetMap database
\footnote{http://www.openstreetmap.org} to perform geo-localized
queries.
The open license ODbL of the database allows to download regional
extracts that can then be loaded in a GIS database (Postgresql+PostGIS
in our case) and queried for a variety of geo-located features.
The data in many urban areas is extremely fine grained, to the level
of buildings and trees.
Furthermore there is a great variety of mapped objects produced by
more that 2 millions users.

In order to extract the quality of a cell $q(x)$, we perform
several queries reflecting different privacy properties
and we combine them in one aggregate number using different weights.
In our experiments we query for a variety of Points Of Interest in the
tag class \texttt{amenity}, such as restaurants and shops, and for the
number of buildings in a cell.
The buildings are an indication of the population density, in fact
despite the database provides a \texttt{population} tag, it is for
large census areas and with a scarce global coverage, while the
\texttt{building} tag can be found everywhere and with fine
resolution.
Considering the simple nature of the queries performed we believe the
resulting grid captures very well the concept of hiding in the crowd
that we wanted to achieve (a sample can be viewed in
Figure~\ref{fig:pp-map}).
We leave more complex query schemes as future work as the main focus
of this paper is on the metric costruction, described in detail in
Section~\ref{algo}.

Among possible improvements three directions seem promising.
% Right now we focus on POIs for the city and everything outside is
% considered open space while in practice many feature of the land may
% be informative.
% Consider the feeling of privacy we might have in a thick wood as
% opposed to an open field, or again the difference between land and
% waterways.
% OpenStreetMap provides numerous tags relative to land features and we
% plan to explore their effectiveness for privacy purposes.

First, one strength of \priv{\dx} is that it is independent of prior
knowledge that an attacker might have about the user, making the
definition suitable for a variety of users.
However in some cases we might want to tailor our mechanism to a
specific group of users, to increase the performance in terms of both
privacy and utility.
In this case, given a prior probability distribution over the
grid of locations, we can use it to influence the privacy mass of each
cell.
For instance, if we know that our users never cross some locations or certain
kind of POIs, we can reduce their privacy mass.

Second, we are interested in queries that reward variety other that
richness e.g. a location with 50 restaurants should be considered less
private than one with 25 restaurant and 25 shops.
Similar ideas have been explored in $l$-diversity, for a detailed
discussion of the differences in our approaches refer to
Section~\ref{sec:related-work}.

Finally, different grids could be computed
for certain periods of the day or of the year.
For instance, our user could use the map described above during the day,
feeling private in a road with shops, but in the evening only a subset
of the tags should be used as many activities are closed, making a road
with many restaurants a much better choice.
% The same could be applied to seasons, imagine for example how snow
% affects human activities in many regions.

Once we have enriched every location $x$ with a quality $q(x)$, we can
compute the resource function $\pp(x)$ as described previously.
In the next part we describe how to exploit this rich and customizable
information to automatically build an elastic metric satisfying a
requirement $\req$.

\subsection{An efficient algorithm to build elastic metrics}\label{algo}
In this section we develop an efficient algorithm to compute a
distinguishability metric $\dx$ that satisfies the quadratic requirement
defined before.
The metric we produce is induced by an undirected graph
$\calg=(\calx,E)$, that is the main structure manipulated by the
algorithm, where vertices are locations and edges $(x,d,x')$ are
labeled with the distance between locations.
The distance between two locations is the shortest path between them
and thanks to this property instead of computing $|\calx|^2$ edges, we
can actually keep just a subset and derive all other distances as
shortest paths.

We start with a fully disconnected graph where all distances are
infinite (thus each location is completely distinguishable) and we
start adding edges guided by the requirement function.
% As a first approach we could work on a single node, adding edges to it
% until its requirement is satisfied and then move to the next node but
% because of the transitivity of the connections we create, this would
% lead to a far too large number of edges.
We work in \emph{iterations} over the grid, where at each iteration we add only one
edge per vertex, stopping when $\req$ is satisfied for all vertices.
The reason to work in iterations is that even if at iteration $i$ a vertex can only reach
a certain number of cells, because of the other edges added during the
same iteration, at $i+1$ it will find itself connected to many more
vertices.
This approach distributes edges uniformly which provides two main
advantages.
First, it increases the locality of connections which in turn reduces the
average error (or increases the utility) of the resulting mechanism.
Second, it leads to a smaller number of edges, thus decreasing the
size of the graph.

The requirement function $\req(l)$ is used as a guideline to define the edges.
Let $\req^{-1}(m) = l^* \sqrt{m}$ be the inverse of $\req$.\footnote{%
	Note that our algorithm is not tied to the specific quadratic requirement
	function, it can work with an arbitrary function $\req$. Even if $\req(l)$ is not
	invertible (e.g. for a step-like requirement), we could use
	$\req^{*}(m) = \inf \{ l \;|\; \req(l) \geq m \}$ in place of $\req^{-1}$.
} This function tells us at what distinguishability level
$l=\req^{-1}(m)$ we should find $m$ amount of privacy mass in order to satisfy the
requirement.
For each location $x$ we keep a temporary level $l_{x}$ that is
updated at each iteration using $\req^{-1}$ and stops at a predefined
maximum value $l^{\top}$.
At the beginning $l_x$ is set using only the privacy mass provided
by $x$ alone but adding edges will take into account also the ball of
points reachable within $l_{x}$.
In other words the temporary level of each location indicates up to
what level of distinguishability the requirement is satisfied.

We then start the iterations and for each vertex $x$ that hasn't already
reached $l^{\top}$ we recompute an updated $l_x$.
The update is necessary to take into account other connections that
may have been added for other vertices and that could increase the
ball of $x$.
In order to add a new edge we need a strategy to find a candidate vertex $x'$ to
connect to.
The strategy we employ is \texttt{next-by-geodistance}, that returns
the cell $x'$ geographically closest to $x$, but still not visited.
In the resulting metric locations are more indistinguishable to nearby
locations, reducing the average error of reported points.
% The candidate vertex is selected by the function
% \texttt{next-by-geodistance} that generates a ring of vertices at the
% same distance from $x$ and iterates inside the ring, when a rings is
% finished the outer one is generated and so on.
% In order to avoid connecting always in the same area, the nodes inside
% a ring are not selected continuously but performing jumps, the length
% of these jumps is a quarter of the ring length so to connect in the 4
% directions uniformly.
Once we have a candidate location $x'$ there are two possible
situations.
If the distance between $x$ and $x'$ is greater than $l_x$, we need to
lower it to satisfy the requirement, so we add an edge $(x,l_x,x')$.
Otherwise if the distance is shorter or equal than $l_x$, this means
that $x'$ is already in the $l_x$ ball of $x$, so we ask
\texttt{next-by-geodistance} for another candidate.
For each vertex not completed an edge is added to the graph and the
process is repeated in iterations until all locations reach $l^{\top}$.

% The intuition behind the algorithm is that when we manage to collect a
% certain amount of privacy point, and therefore to reach a certain
% distance $r_x$, in order to reach slightly further we have no other
% choice but to connect a new location.

% The intuition behind the algorithm is that by connecting a new node at
% a distance that we can already reach with the privacy points we have,
% following the requirement, we obtain two effects.
% First we increase the ball and thus force the algorithm to proceed.
% Second we make sure that we are always slightly exceeding the
% requirement.

Our experiments showed that completion of the last few tens of
vertices can take extremely long and they are localized mostly on the
border of the grid.
This is due to the fact that points close to the border have
fewer neighbors, making it harder for them to find a candidate to connect to.
As a consequence they need to reach much further away, taking more
iterations and resulting in a higher average error because of the long
connections created.
For this reason we use a stopping condition that checks, at the end of
every iteration, if all the nodes remaining to complete are closer to
the border than a certain \emph{frame} constant.
If they are, the algorithm stops without completing their requirement.
All locations inside this frame of the grid can be reported as
sanitized locations but cannot be used as secret locations.
The \emph{frame} value is a compromise between the algorithm running
time and usable grid size, in our experiments we used 3\% of the grid
size.

It can be shown that the metric $\dx$ constructed by this algorithm does satisfy
the requirement $\req$ for all $l\le l^\top$. The stopping level $l^\top$ can be
set arbitrarily high, but in practice setting it to any value larger than 10 has
no effect on the resulting metric. As shown in the evaluation of
Section~\ref{sec:evaluation}, the algorithm can scale to an area of half a
million locations with modest computing resources. This is several orders of
magnitude better that techniques computing optimal obfuscation mechanisms
\cite{Shokri:12:CCS,Bordenabe:14:CCS,Shokri:14:TechRep} which can only handle a
few hundred points within reasonable time constraints. Of course, our method
gives no optimality guarantees, it only constructs one possible metric among
those satisfying the requirement.

We believe further improvements in performance are possible in three
directions, that we leave as future work.
When working in privacy poor areas, like in the country, the algorithm
spends a considerable time compressing large areas, as expected.
We believe that this work could be avoided by grouping together
several locations already when laying down the grid.
We would have a coarser resolution in the country, which is
acceptable, and a large speed up in the metric construction.
A second obvious improvement would come from running the algorithm in
parallel on portions of the map and merging the results.
The problem arises on the borders of the submaps, where on adjacent
locations we have connections with sharply different shapes.
We believe however that computing several submaps leaving a frame of
uncompleted points and them running again the algorithm in the entire
map could provide a reasonable result.
Finally several strategies can be applied in the choice of the next
candidate and in the way we perform the iterations that could have an
impact on speed and utility of the mechanism by completing faster the
requirement.

\begin{figure}[t]
\begin{lstlisting}[mathescape,language=java, emph={foreach},emphstyle=\textbf]
  foreach $x \in \calx$ do
    $l_{x}$ := $\req^{-1}(\pp(\{x\}))$
  while $\exists x.\; l_x \neq l^{\top}$ do
    foreach $x \in \calx$ do
      $l := \req^{-1}(\pp(B_{l_x}(x)))$
      do 
        $x' := \texttt{next-by-geodistance}(x)$
      while $\dx(x,x') \leq l_x$
      $E := E \cup \{(x,l_x,x')\}$
\end{lstlisting}
\end{figure}

\subsection{Practical considerations}
We believe that the techniques presented are practical enough to
deploy a location privacy mechanism in a real setting.
The resources required both in terms of hardware and time are very
limited, consider that the mechanism evaluated in the next section was
built in a day on a medium Amazon E2C instance.
As already mentioned querying the database is easily parallelizable
and the same grid can be reused to build several metrics.
We could imagine having a choice of pre-computed grids, for different
flavors of location quality and times (as explained in Sec~\ref{queries}), on
top of which the user could tune the requirement.

The computation of the metric is the most demanding part of the
process but proved reasonably fast for an area that can easily contain
all the movements of the user and many optimization are still possible
in the algorithm.
% If the user does not apply any fence, as for the grid, he could choose
% from a set pre-computing metrics avoiding the delay of the
% re-computation.
We imagine these two computationally intensive activities to be
performed by a remote server, with seldom updates from the OpenStreetMap
database.

The user's phone needs to take care only of downloading an extract of the
metric to use in the Exponential mechanism.
For every request, the mechanism computes the exponential distribution
of sanitized locations and draws from it, which amounts to a trivial
computation, both in memory and time.
In principle we could also avoid contacting a third party by saving
the entire metric locally on the phone, as a reference our metric for
Il\^e de France is only 58MB.

From a user's perspective, the amount of configuration required to run
the mechanism varies according to her needs.
It can go from no configuration, in the case she downloads a
pre-computed map, to scaling the privacy mass requirement, to full
customization in the case the users wants to tailor the queries to her
specific needs.

%%% Local Variables: 
%%% mode: latex
%%% TeX-master: "paper"
%%% End: 

\section{Incorporating fences in the metric}
\label{sec:fenced_metric}

As discussed in the introduction another issue of
\geoind{} is that, repetitive use of a mechanism from
the same location is bound to reveal that location as the number of
reports increases.
This is crucial for locations that the user frequently visits, such as
his home or work location.
Such locations cannot be expected to remain indistinguishable in the long run; repetitive
use of the mechanism is bound to reveal them with arbitrary accuracy.

Despite the fact that all privacy mechanism are susceptible to this privacy erosion
over time, the compositionality property of \egeoind{} quantifies exactly this
privacy degradation:
repetitive use of a mechanism causes a linear
accumulation of $\epsilon$, lowering the privacy protection
guaranteed.
% In practice using a geo-indistinguishable mechanism in such a location
% would lead to a cloud of reported locations perfectly centered on the
% secret location, that would be identifiable with great accuracy.
There are techniques to alleviate this effect, such as the predictive
mechanism of \cite{Chatzikokolakis:14:PETS}, but they are not enough
in this highly recurrent cases.

This problem, especially for the home-work locations, has been already
studied in the literature \cite{Golle:09:PerCom}, although in the
context of anonymity i.e. how to match a user identity to a pair of
home-work locations.
In fact our interest is focused on reducing the accuracy of the
reported location, because of the sensitive data the attacker can
infer from it.
Even if the user is authenticated with a LBS, for instance to notify her
friends that she is home, it is still valuable to not disclose the
precise address (that friends know anyway).

% Furthermore work locations are often already partially disclosed. 
% For example many institutions provide a webpage of each employee with
% a more or less precise address.
% Professions in contact with the public, such as shops or health
% facilities, necessarily publish their location information.
% Our intuition is that highly recurrent locations need a separate
% treatment and while they can't be indistinguishable from any another it
% is important to reduce their local accuracy.

\subsection{Fences}
For highly recurrent locations we propose the use of geo
\emph{fences}, areas on the map where the user's movements are
completely hidden and that are considered known to the attacker.
This technique is not novel and indeed has been widely used by a
large number of LBS.
For example, personal rental services (e.g. Airbnb) allow the user to
indicate an area where the good to be rented is located, so that other
users can evaluate if it is at a convenient distance without
compromising the privacy of the owner.
Despite the vast use of fences in practice, to the best of our
knowledge, there is a lack of works in the literature about their
implementations or evaluating their effectiveness.

Our contribution consist in a simple formalization of fences in the
framework of distinguishability metrics.
This construction allows to hide completely sensitive locations
within a fence, while permitting the use of any other $\dx$-private
mechanism outside.
Given a privacy metric $\dx$, we define a new fenced metric $d_F$ as:
\[ d_F(x,x') = 
\left\{
\begin{array}{lcl}
  0      & x,x' \in F \\
  \dx(x,x') & x,x' \notin F \\
  \infty & \text{otherwise}
\end{array}
\right.\]
Outside the fence the original metric $\dx$ is in place while inside the
fence all points are completely indistinguishable.
The advantage is that being zero the distance \emph{inside} the fence,
any repeated use of the mechanism from the sensitive location comes
for \emph{free}, effectively stopping the linear growth of the budget.
In practice we keep reporting uniformly points inside the fence, thus
leaking no information to the attacker, other than the public fact
that we are inside the fence.
Indeed it should be noted that any movement \emph{across} the fence is
completely distinguishable, the attacker knows when we are in or out.

Regarding utility, in this case it simply depends on the size of the
fence, in direct contrast with privacy.

\paragraph*{Automatic Configuration of position and size}
In order to configure the position and size of the fences, the user
input would be the best option (as shown in \cite{Brush:10:UbiComp}),
however they could also be inferred and suggested automatically.
In \cite{DBLP:journals/tdp/GambsKC11} the authors developed an attack
to identify POI of a specific user, from a set of mobility traces.
A similar technique could be employed on the user's phone, over a
training period, to collect and analyze her movements for a few days.
The mechanism would then automatically detect recurrent locations and
suggest the user to fence them, possibly detecting more than just
home/work locations.

With the use of geolocated queries, such as those used to extract
privacy points in Section~\ref{queries}, we could determine the size
of the fence so to include a reasonable amount of buildings for home
and other POIs for work.

\paragraph*{Compatibility with the elastic metric algorithm}
It should be noted that fences are applicable to any
distinguishability metric, including but not limited to the elastic
metric presented in Sec~\ref{sec:elastic_metric}.
Not only we can incorporate fences in our elastic definition but also
in our graph based algorithm, in a simple and efficient way.
It amounts to connecting the locations inside the fence with zero
labeled edges and to leave them disconnected from the nodes outside.
When running the algorithm we should also take care to maintain this
disconnection of the fence.
In order to avoid adding edges from the inside we simply set the
temporary radius $r_x$ of all nodes inside the fence to $d^{\top}$, so
that the algorithm considers them completed and skips them.
On the other side, to avoid adding edges from the outside, we need to
modify the function \texttt{next-by-geodistance} so to avoid
considering a candidate any location inside a fence.
Both alterations to the algorithm are trivial to implement and have no
effect on performances.
The only drawback of the presented method is that the fences need to
be set before building the metric, which is inconvenient as for each
user we are obliged to recompute, for the most part, the same
metric.
Despite this our experiments show that in under a day is possible to
generate the metric so this remains an effective technique for
practical purposes, especially considering the very static nature of
the fences.

As future work we are investigating the possibility to carve the
fences from an already built metric.
The challenge in this case is re-enforcing the requirement for all
affected points and it is not clear for now how spread out is the
impact on other nodes and what is the effect on utility.

%%% Local Variables: 
%%% mode: latex
%%% TeX-master: "paper"
%%% End: 

\section{Evaluation}
\label{sec:evaluation}

\begin{figure}[t]
  \centering
  \includegraphics[width=\columnwidth]{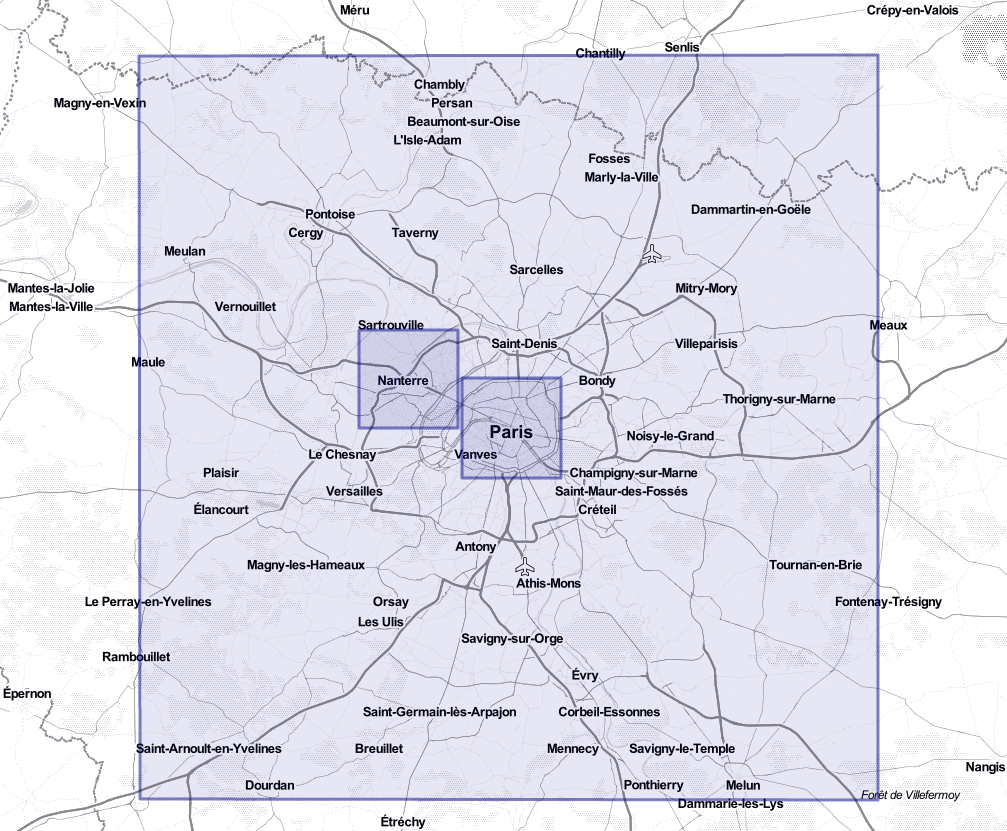}
  \caption{Coverage of \emm{} with two subregions: Nanterre suburb on the left and Paris city on the right}
  \label{fig:coverage}
\end{figure}

In this section we perform an extensive evaluation of our technique in Paris'
wide metropolitan area, using two real-world datasets. We start with a
description of the metric-construction procedure, and we discuss the features
of the resulting metric as well as the obfuscation mechanism obtained from it.
Then, we compare the elastic mechanism to the Planar Laplace mechanism
satisfying \geoind{}, using data from the Gowalla and Brightkite social
networks. The comparison is done using the privacy and utility metrics of
\cite{Shokri:11:SP}. It should be emphasized that the metric construction was
completely independent from the two datasets, which were used only for the
evaluation.
All the code used to run the evaluation is publicly available \cite{elastic-github}.

\subsection{Metric construction for Paris' metropolitan area}

\begin{figure*}[t]
  \centering
  \subfloat[Privacy mass $m(x)$ of each location]{
    \includegraphics[width=0.445\textwidth]{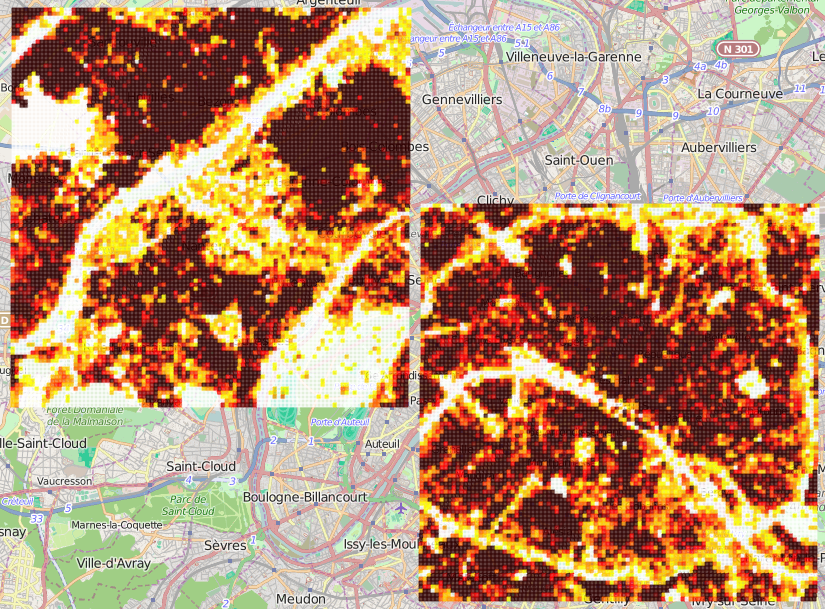}
    \label{fig:pp-map}}
  \subfloat[Expected error $E_\mathemm(x)$ at each location]{
    \includegraphics[width=0.45\textwidth]{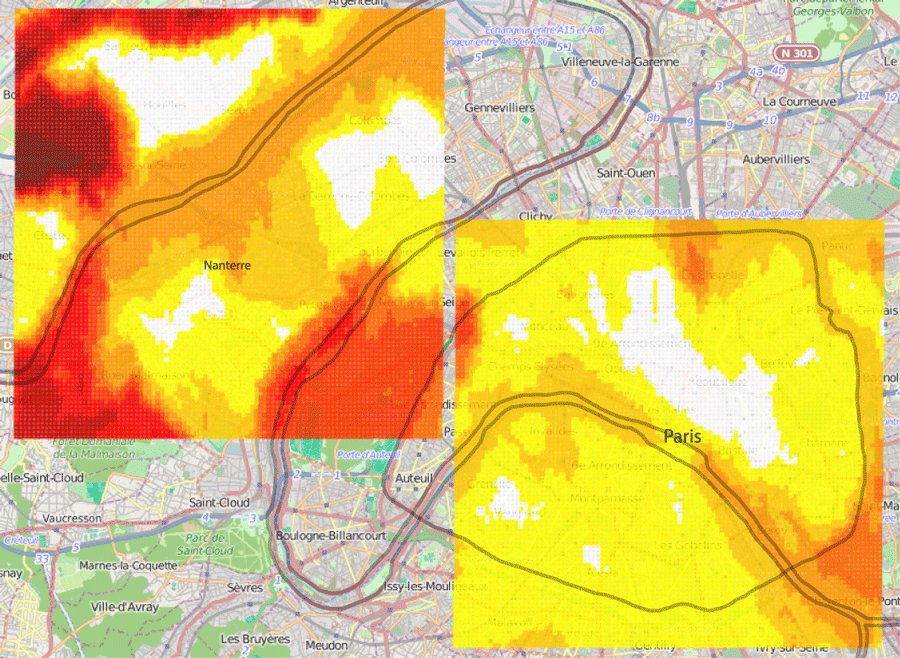}
    \label{fig:em-err}}
  \caption{Paris' center (right) and the nearby suburb of Nanterre (left)}
\end{figure*}

We build an elastic metric $\dx$ for a $75$ km $\times$ $75$ km
grid centered in Paris, roughly covering its extended metropolitan area.
Each cell is of size $100$ m $\times$ $100$ m, and the set of locations $\calx$
contains the center of each cell, giving a total number of $562,500$
locations. The area covered is shown in Fig~\ref{fig:coverage}; note that
the constructed metric covers the larger shown area, the two smaller ones are
only used for the evaluation of the mechanism in the next section.

The semantic quality $q(x)$ of each location was extracted from OpenStreetMap
as explained in Section~\ref{queries}, and the privacy mass $m(x)$ was computed
from \eqref{eq:resource} using $\rcity = 300$ m and $\rcountry = 3$~km. The
resulting mass of each location is shown in Figure~\ref{fig:pp-map}, where white
color indicates a small mass while yellow, red and black indicate increasingly
greater mass. The figure is just a small extract of the whole grid depicting the
two smaller areas used in the evaluation: central Paris and the nearby suburb of
Nanterre. Note that the colors alone depict a fairly clear picture of the city:
in white we can see the river traversing horizontally, the main ring-road and
several spots mark parks and gardens. In yellow colors we find low density areas
as well as roads and railways while red colors are present in residential areas.
Finally dark colors indicate densely populated areas with presence of POIs.

For this grid, we use the algorithm presented in Section~\ref{algo} to compute
an elastic metric $\dx$ with the quadratic requirement of
\eqref{eq:requirement}, configured with $l^* = \ln2$. The whole computation took
less than a day on an entry-level Amazon EC2 instance. This performance of the
algorithm is already sufficient for real-world use: the metric only need to be
computed once, while the computation can be done by a server and the result can
be then transmitted to the user's device. Note that the algorithm can deal with
sizes several orders of magnitude bigger that techniques computing optimal
obfuscation mechanisms \cite{Shokri:12:CCS,Bordenabe:14:CCS,Shokri:14:TechRep},
which makes it applicable to more realistic scenarios.

We then construct an Exponential mechanism (described in
Section~\ref{sec:preliminaries}) using $\dx$ as the underlying metric. We refer
to the resulting obfuscation mechanism as the Elastic Mechanism (\emm). The
mechanism is highly adaptive to the properties of each location: high-density
areas require less noise than low-density ones to achieve the same privacy
requirement. Figure~\ref{fig:em-err} shows our utility metric per location,
computed as the expected distance $E_\mathemm(x)$ between the real and the
reported location. Compared to Figure~\ref{fig:pp-map} it is clear that areas
with higher privacy mass result to less noise. Populated areas present a good
and uniform error that starts to increase on the river and ring-road. On the
other hand, the large low-density areas, especially in the Nanterre suburb, have
a higher error because they need to report over larger areas to reach the needed
amount of privacy mass. 

\begin{figure}[t]
\centering
  \includegraphics[width=0.7\columnwidth]{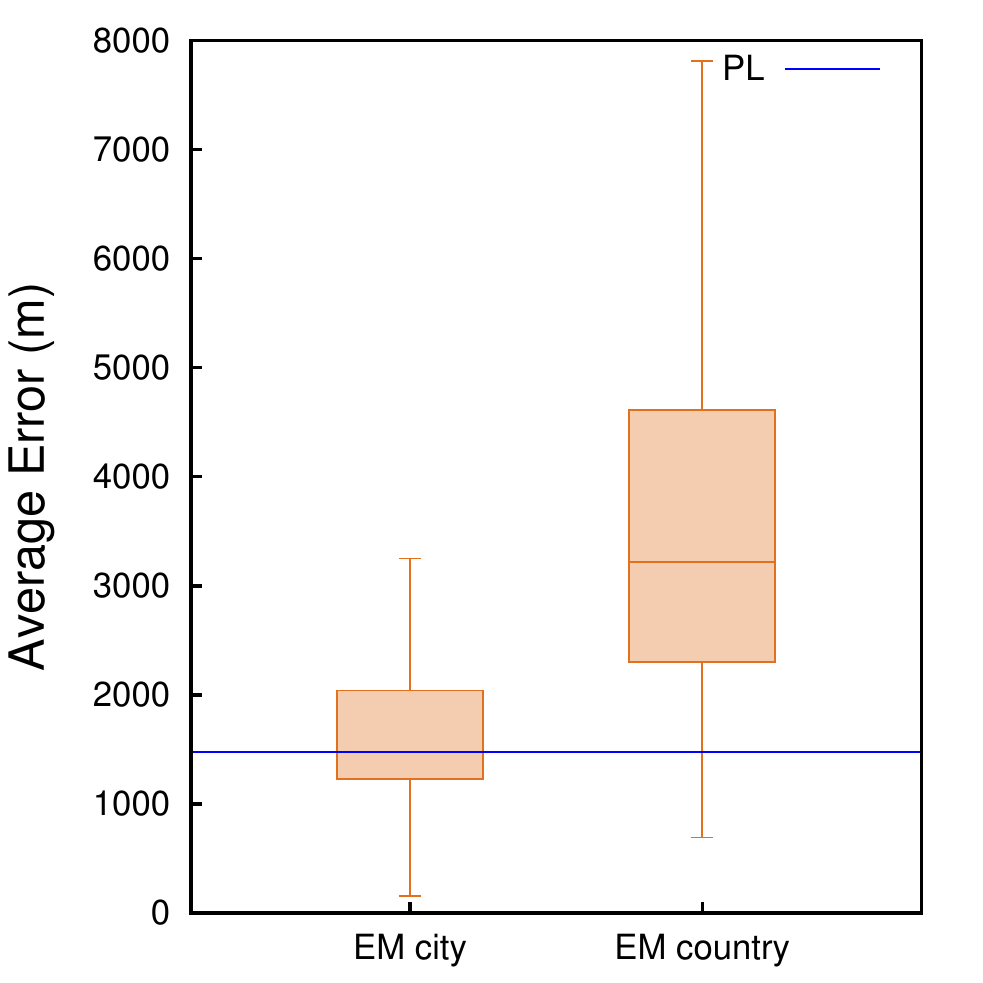}
  \caption{Expected error $E_\mathemm(x)$ per location}
  \label{fig:utility-users}
\end{figure}

Finally, Figure~\ref{fig:utility-users} shows a boxplot of the expected error
for each location in the two areas. It is clear that the amount of noise varies
considerably, ranging from a few hundred meters to several kilometers. It is
also clear that locations in central Paris need considerably less noise that
those in the suburban area. For comparison, the Planar Laplace mechanism
(compared against \emm{} in the next section) has a constant expected error for
all locations.

Note that the expected error will always be higher than the
$\rcity$ used in the normalization. For example in a location that
satisfies its requirement in $300$ m it would be $870$ m.  This is
expected and it is due to the nature of the exponential noise
added.

\subsection{Evaluation using the Gowalla and Brightkite datasets}

In this section we compare the Elastic Mechanism (\emm) constructed in the
previous section with the Planar Laplace mechanism \cite{Andres:13:CCS}
satisfying standard \geoind. For the evaluation we use two real-world datasets
from location-based social networks.

\paragraph{The Gowalla and Brightkite datasets}

Gowalla was a location-based social network launched in 2007 and closed in 2012,
after being acquired by Facebook. Users were able to ``check-in'' at locations
in their vicinity, and their friends in the network could see their check-ins.
The Gowalla dataset \cite{Cho:11:SIGKKD} contains $6,442,890$ public check-ins
from $196,591$ users in the period from February 2009 to October 2010.
Of those, $9,635$ check-ins were made in Paris' center area and $429$ in the
Nanterre area (displayed in Fig~\ref{fig:coverage}).

Brightkite was another location-based social network created in 2007 and
discontinued in 2012. Similarly to Gowalla users could check-in in nearby
locations and query who is nearby and who has been in that location before.
The Brightkite dataset \cite{Cho:11:SIGKKD} contains $4,491,143$ check-ins
from $58,228$ users. Of those, $4,014$  check-ins were made in Paris' center
and $386$ in Nanterre.

These datasets are particularly appealing for our evaluation since a check-in
denotes a location of particular interest to the user, and in which the user
decided to interact with an actual LBS. This is in sharp contrast to datasets
containing simply mobility traces, which just contain user movements without
any information about the actual use of an LBS.

\paragraph{Privacy metrics}

\begin{figure*}[t]
  \centering 
  \subfloat[Gowalla]{
    \includegraphics[width=0.5\textwidth]{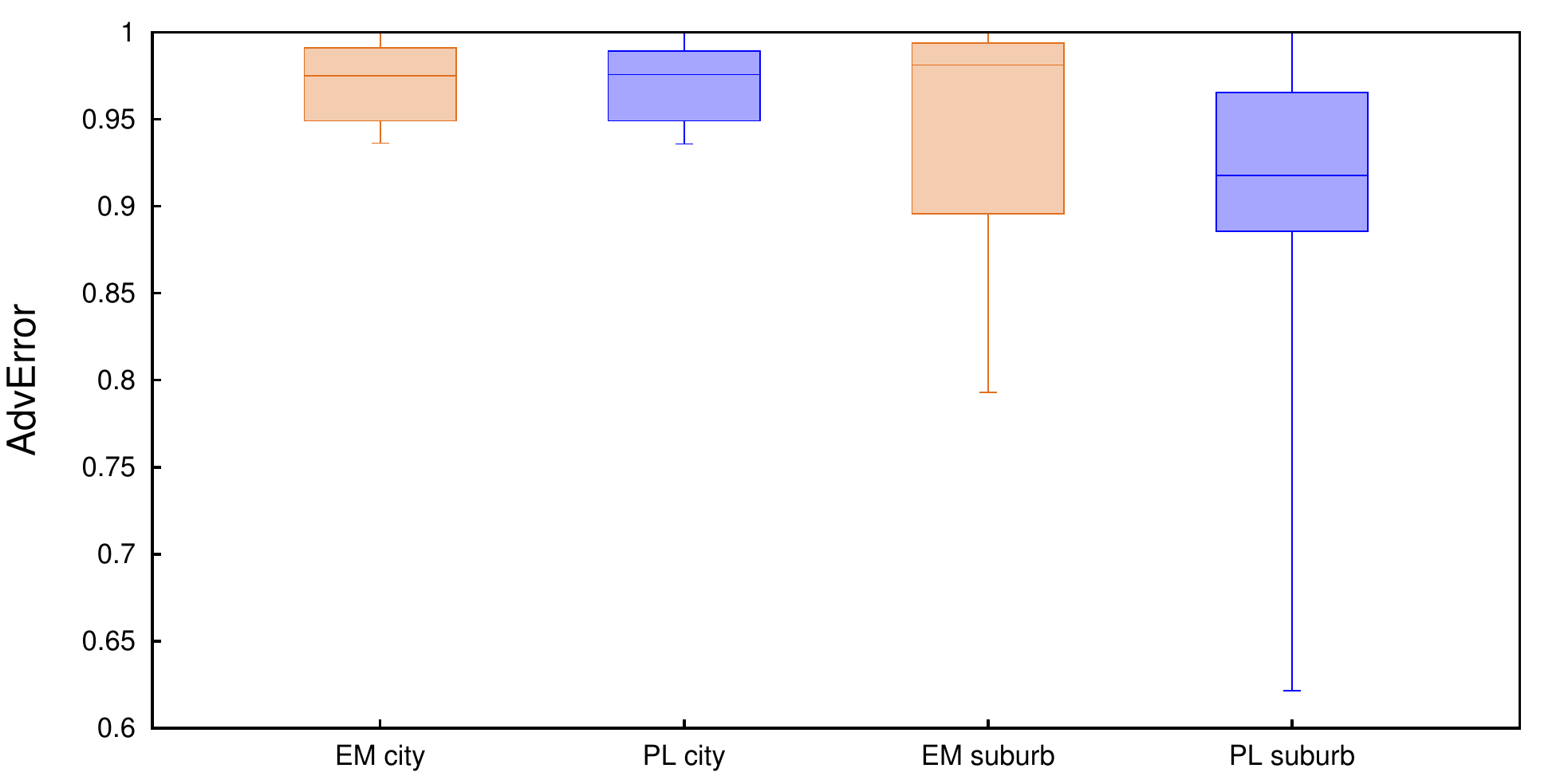}
  }
  \subfloat[Brightkite]{
    \includegraphics[width=0.5\textwidth]{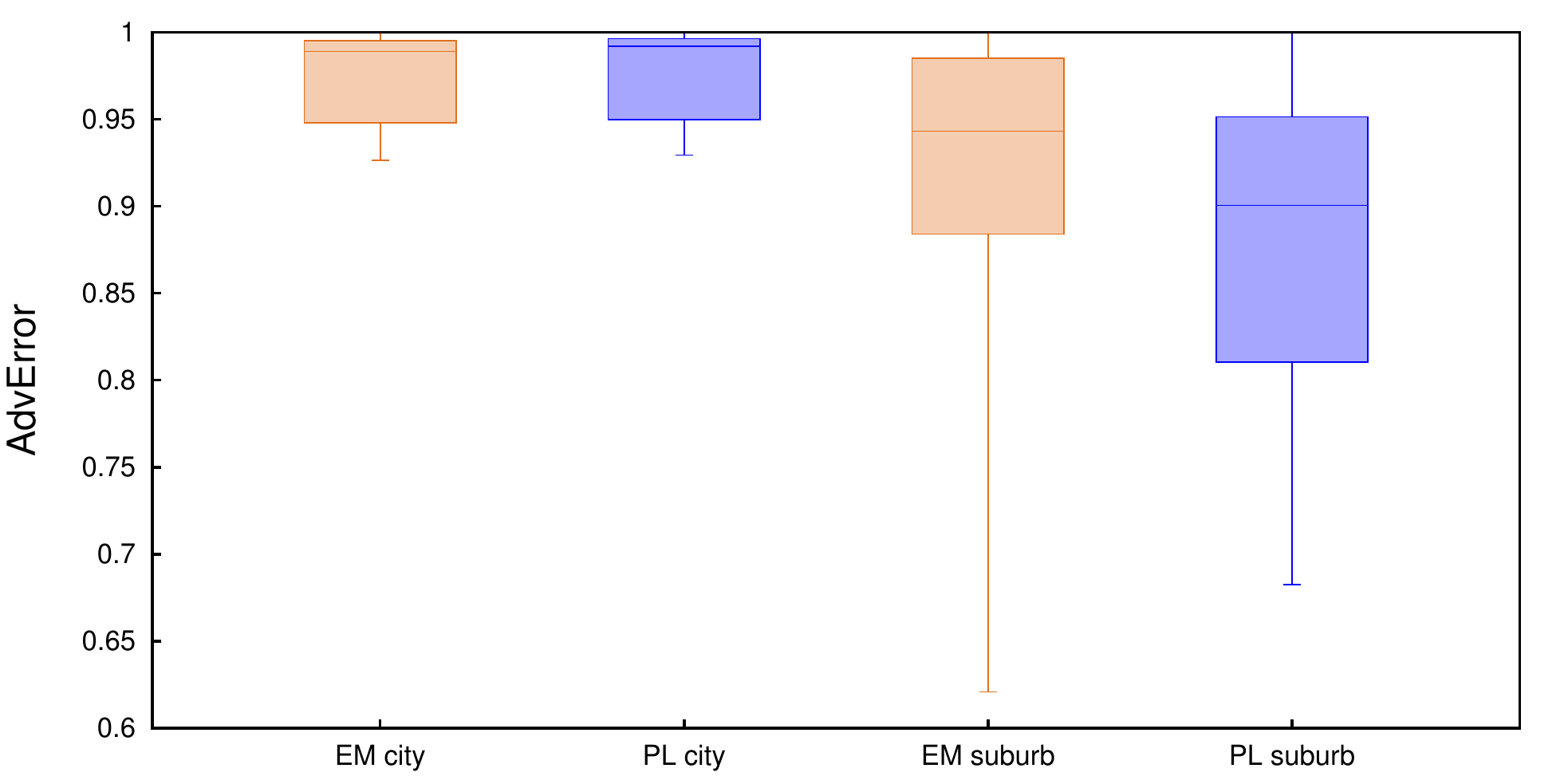}
  }
  \caption{Per-user binary $\adv$ of the \emm{} and \pl{} mechanisms for each area}
  \label{fig:adverr-binary}
\end{figure*}

Since the \emm{} and \pl{} mechanisms satisfy different privacy definition, to
perform a fair comparison we employ the widely used Bayesian privacy metric of
 \cite{Shokri:11:SP}.
This metric considers a Bayesian adversary having a prior knowledge $\pi$ of the
possible location of the user and observing the output of the
mechanism $K$.
After seeing a reported location $z$ the attacker applies a strategy
$h: \calz \to \calx$ to remap $z$ to the real secret location where he
believes the user could be, e.g. if $z$ is in a river, it is likely
that the user is actually in a nearby location $x$ on the
banks.

The \adv{} metric measures the expected loss of an attacker trying to infer the
user's location. It is defined as:
\[
\advarg{K}{\pi}{h}{\da} = \sum_{x,z} \pi(x) \; K(x)(z) \; \da(x,h(z))
\]
where $\da$ is a metric modeling the adversary's loss in case he fails to
identify the user's real location. Notice how the loss function is not applied
directly to the reported location $z$, but to the remapped location $h(z)$.
A rational adversary will use the strategy $h^*$ minimizing his error, hence
\cite{Shokri:11:SP} proposes to use $\advarg{K}{\pi}{h^*}{\da}$ as a privacy
metric, where
\[ h^{*} = \argmin_h \; \advarg{K}{\pi}{h}{\da} \]
Note that $h^*$ can be computed efficiently using the techniques
described in \cite{Shokri:12:CCS}.

In our evaluation the secrets are POIs in each dataset.
We use two commonly used loss functions modelling different
types of adversaries:
the first is the \emph{binary} loss function $\binary$,
modelling an adversary interested in the semantics of the user's POI,
hence trying to guess exactly in which one he is located.
\[
\binary(x,z) = \left\{
\begin{array}{ll}
  0 & x = z \\
  1 & x \neq z
\end{array}
\right.
\]
Then $\advarg{K}{\pi}{h}{\binary}$ expresses the
adversary's \emph{probability of error} in guessing the user's POI. Second, we
use the Euclidean loss function $\euclid$, modelling an adversary who is interested in
guessing a POI \emph{close} to the user's, even if that POI is unrelated to his
activity. In this case, $\advarg{K}{\pi}{h}{\euclid}$ gives the adversary's
\emph{expected error} in meters in guessing the user's POI.

We should emphasize an important difference between the two adversaries:
$\binary$ tries to extract semantic information from the actual POI, and is less
effective in dense areas where the number of POIs is high. On the other hand,
$\euclid$ is less sensitive to the number of POIs: if many POIs are close to
each other, guessing any of them is equally good. This difference is clearly
visible in the evaluation results.

We use each dataset to obtain a prior knowledge $\pi^*$ of an ``average''
user of each social network, by considering all check-ins within the areas
of interest. Note that the datasets do not have enough check-ins \emph{per-user}
to construct \emph{individual} user profiles. Indeed, most users have checked-in
in each location at most once, hence any profile built by $n-1$ check-ins would
be completely inadequate for inferring the remaining one. As a consequence, we
assume that the adversary will compute his best strategy $h^*$ by using the
global profile $\pi^*$.

Finally, we use
	\[ 
	\advarg{K}{\pi_u}{h^{*}}{d_{A}}
	\qquad d_A \in \{\binary,\euclid\}
\]
as our privacy metric, where $\pi_u$ is the user's individual prior
(computed only from the user's check-ins), and
\[ h^{*} = \argmin_h \; \advarg{K}{\pi^{*}}{h}{d_{A}} \]
is the adversary's strategy computed from the global profile. Hence, the
individual priors $\pi_u$ are only used for averaging and not for constructing
the strategy.

\paragraph{Utility metrics}
The utility of an obfuscation mechanism is in general closely tied to the application at hand.
In our evaluation we want to avoid restricting to a particular one; as a
consequence we use two generic utility metrics that are reasonable for a variety
of use cases. We measure utility as the \emph{expected distance} between the real
and the reported locations, using two distance functions.
In some applications, service quality degrades linearly as the reported
location moves away from the real one; in such cases the Euclidean distance
$\euclid$ provides a reasonable way of measuring utility. Other applications
tolerate a noise up to a certain threshold $r$ with almost no
effect on the service, but the quality drops sharply after this threshold. In
such cases, we can measure utility using the follwoing distance metric:
\[
\threshold(x,z) = \left\{
\begin{array}{ll}
  0 & \euclid(x,z)<r\\
  1 & ow.
\end{array}
\right.
\]

\paragraph{Results}

\begin{figure*}[t]
  \centering 
  \subfloat[Gowalla]{
    \includegraphics[width=0.5\textwidth]{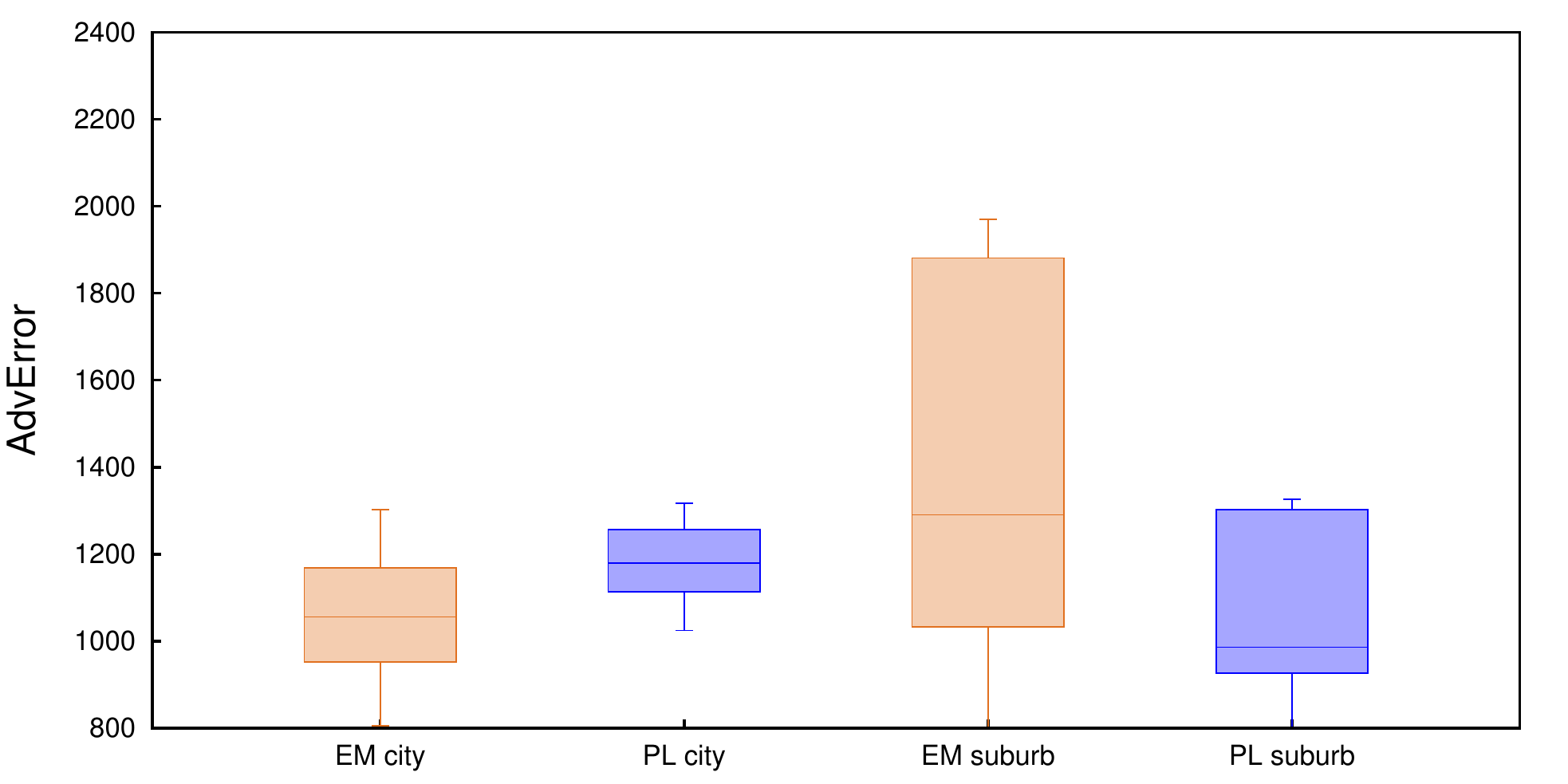}
  }
  \subfloat[Brightkite]{
    \includegraphics[width=0.5\textwidth]{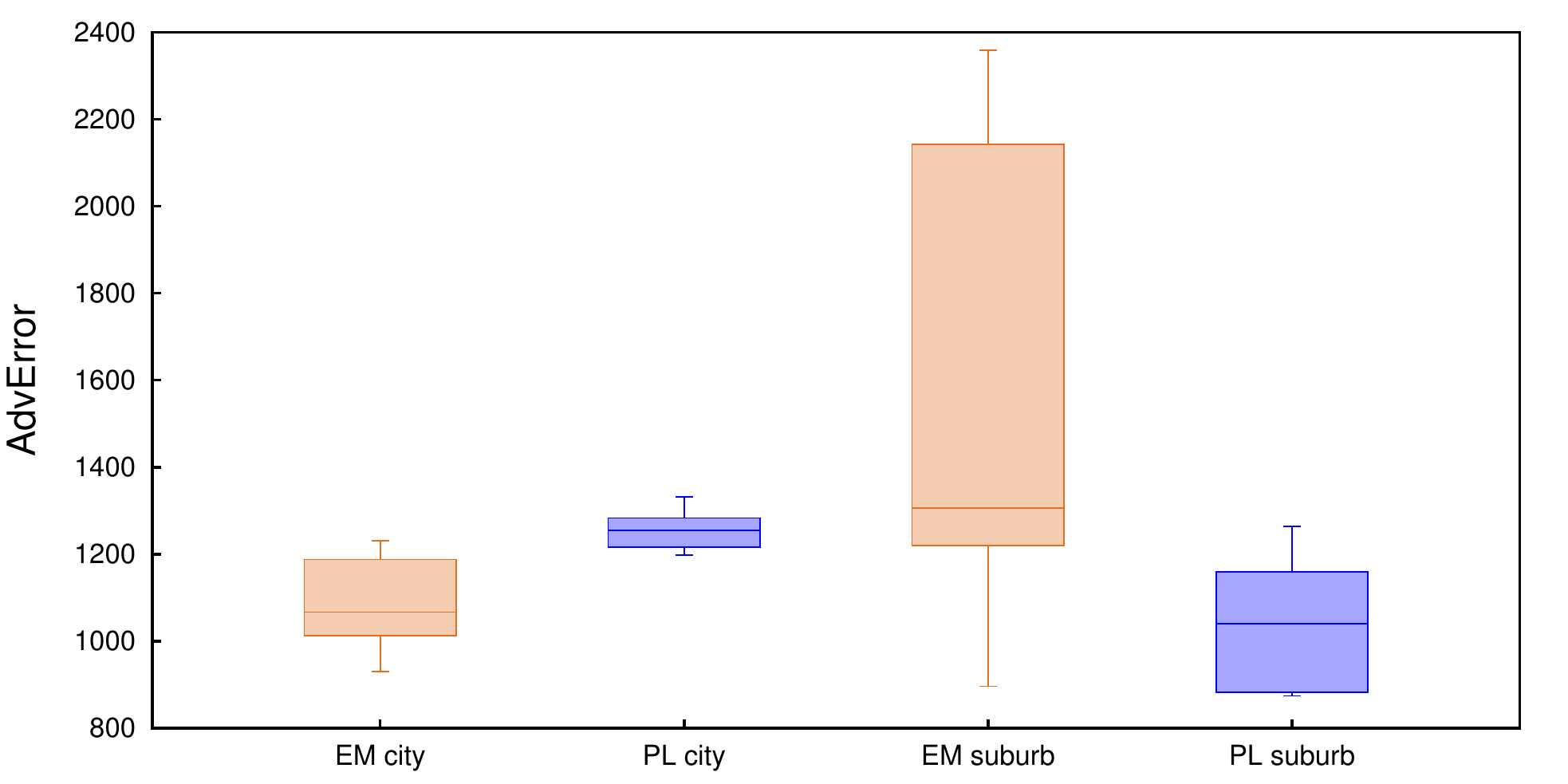}
  }
  \caption{Per-user Euclidean $\adv$ of the \emm{} and \pl{} mechanisms for each area}
  \label{fig:adverr-euclidean}
\end{figure*}

We carry out our evaluation in two different areas of the Paris metropolitan
area, with very different privacy profiles. The first area is the city of Paris
intra-muros, very private on average, while the second is the adjacent suburb of
Nanterre, where already the concentration of privacy mass is much lower.

To obtain a fair comparison, the Planar Laplace \pl{} mechanism is configured to
have the same utility as the \emm{} mechanism in Paris' center. We computed the
utility of \emm{} using the global profile $\pi^*$ from both datasets, using
four distance functions, namely $\euclid$ and $\threshold$ with $r = 1200, 1500$
and $1800$ meters. In each case, we computed the parameter $\epsilon$ of \pl{}
that gives the same utility; we found that in all 8 cases the $\epsilon$ we
obtained was almost the same, ranging from $0.001319$ to $0.001379$. Since the
difference between the values is small, we used a single configuration of \pl{}
with the average of these values, namely $\epsilon = 0.001353$. With this
configuration, we then compare the two mechanisms' privacy in both areas.

Note that, although we configured the \emph{expected} utility of both mechanisms
to be the same in the city, the \emm{}'s behaviour is highly dynamic: in
the most private areas of Paris \emm{} uses much less noise: in 10\% of the
city's locations \pl{} adds 50\% or more noise than \emm{}.

The results for the binary adversarial error are shown in Figure~\ref{fig:adverr-binary}. We can see that in the
center of Paris both mechanisms provide similar privacy guarantees.
The dynamic nature of \pl{} does not affect its privacy: locations
in which the noise is lower have more POIs in proximity, hence even with
less noise it is still hard to guess the actual one.
On the other hand,
in the suburb there is a sharp degradation for \pl. The reason is that the
number of POIs in both datasets is much smaller in Nanterre than in Paris,
the distance between them is greater, and the resulting priors are more
``informed''. Hence it is considerably easier for
the adversary to distinguish such POIs, leading to a probability of error as low
as $0.62$. On the other hand, \emm{} maintains a higher $\adv$ in the suburb, by
introducing a higher amount of noise.

To match \emm{}'s privacy guarantees in the suburb, \pl{} should be configured
with a higher amount of noise. However, since Planar Laplace treats space in the
same way everywhere, this would lead to a high degradation of utility in Paris'
center, which is unfortunate since (i) the extra noise is unnecessary to provide
good privacy in the center, and (ii) the extra noise could render the mechanism
useless in a dense urban environment where accurate information is crucial. In
short, the flexibility of the elastic mechanism allow it to add more noise when
needed, while offering better utility in high-density areas.

The results for the Euclidean adversarial error are shown in
Figure~\ref{fig:adverr-euclidean}. Here, we see a sharp difference wrt the
binary adversary: the effectiveness of guessing a POI \emph{close} to the real
one is not affected much by the number of POIs (guessing any of them is equally
good). As a consequence, \emm{}, which adds less noise in dense areas with a
great number of POIs, scores a lower adversarial error in the city (although the
difference is moderate). The median
error for \emm{} in Gowalla is 1056 meters while for \pl{} it is 1180 meters.
The motivation behind \pl{} was that, in a dense urban area, it is harder for
the adversary to extract semantic information from the reported location even if
less noise is used. But the Euclidean adversary is not interested in semantic
information, so he scores better with less noise.

In the suburb, on the contrary, the picture is reversed. Due to the fact that
priors in Nanterre are more ``informed'' making remapping easier, the
adversarial error for \pl{} decreases compared to the city. On the other hand,
\emm{} adapts its noise to the less dense environment, leading to much higher
adversarial error.

Note that the Nanterre area is still quite close to the city center and highly
populated itself. The fact that we can already see a difference between the two
mechanisms so close to the center is remarkable; clearly, the difference will be
much more striking as we move away from the city center (unfortunately, the
datasets do not contain enough data in these areas for a meaningful quantitative
evaluation).

Finally, we should emphasize that the mechanism's construction and evaluation
were completely independent: no information about Gowalla's or Brightkite's list
of check-ins was used to construct the metric. The only information used to
compute $\dx$ was the semantic information extracted from OpenStreetMap.

%%% Local Variables: 
%%% mode: latex
%%% TeX-master: "paper"
%%% End: 

\section{Related Work and Conclusion}\label{sec:conclusions}
\paragraph{Related work}\label{sec:related-work}
Concerning location privacy, there are excellent works and surveys
\cite{Terrovitis:11:SIGKDD,Krumm:09:PUC,Shin:12:WC} that present the
threats, methods, and guarantees.

A large body of works developed from $k$-anonymity, originally a
notion of privacy for databases
\cite{Samarati:01:TKDE,Kido:05:ICDE,Shankar:09:UbiComp,Bamba:08:WWW,Duckham:05:Pervasive,Xue:09:LoCa,Machanavajjhala:07:TKDD}.
Many of the shortcomings of $k$-anonymity, outlined in
\cite{Shokri:10:WPES}, are addressed in the current main trend based
on the expectation of distance error
\cite{Shokri:11:SP,Shokri:12:CCS,Hoh:05:SecureComm,Dewri:12:TMC}.
Both are dependents on the adversary's side information, contrary to our approach, as are some other
works \cite{Cheng:06:PET} and \cite{Ardagna:07:DAS}.

Extraction of privacy points uses ideas similar to $k$-anonymity or
$l$-diversity but the mechanisms are very different in nature.
The privacy of our resulting mechanism is due to the obfuscation
technique, it is independent of the attacker side knowledge and
doesn't require any third party to operate.
Furthermore we protect privacy as the location accuracy and not its
anonymity.

Notions that abstract from the attacker's knowledge based on
differential privacy can be found in \cite{Machanavajjhala:08:ICDE}
and \cite{Ho:11:GIS} although only for \emph{aggregate} information.

In this work we extend and generalize \geoind{}
\cite{Andres:13:CCS} and in order to so we go back to the
notion it is based on, \priv{\dx} \cite{Chatzikokolakis:13:PETS} that
provides a metric extension of differential privacy.
This family of definitions abstracts from the attacker's prior
knowledge, and is therefore suitable for scenarios where the prior
is unknown, or the same mechanism must be used for multiple users.

Regarding the construction of finite mechanisms,
\cite{Shokri:14:TechRep} proposes a linear programming technique to
construct an optimal obfuscation mechanism with respect to either the
expectation of distance error or \geoind{}.
In \cite{Bordenabe:14:CCS} the authors propose again a linear
programming technique to compute a geo-indistinguishable mechanism
with optimal utility.  Their approach uses a spanner graph to
approximate the metric in a controlled way.
Our algorithm does not provide optimality with respect to privacy nor
utility, it guarantees the respect of a privacy requirement while
achieving good utility.
Moreover the state of the art in optimal mechanism construction is
limited to few tens of locations while the purpose of our technique is
to scale to several thousands of points.

% The predictive mechanism reduces the budget usage by skipping the cost of sanitizing some locations that are easily predictable.
% On the other hand using its prediction function in a private way as a
% (small) cost on the budget that is still linear with the number of
% uses.
% Indeed this cost would accumulate even if the prediction function
% could predict very accurately home/work locations, while our solution
% of putting fences at the metric level doesn't suffer from this.
% Note however the predictive mechanism is still useful in the larger
% set of cases where there is considerable correlation on the user's
% traces and can be used in combination with fences.

\paragraph{Conclusion} In this paper, we have developed a novel elastic privacy
metric that allows to adapt the privacy requirement, as hence the amount of
applied noise, to the properties of each location. We have formalized a
requirement for such metrics based on the concept of privacy mass, and using
semantic information extracted from OpenStreetMap. We have developed a
graph-based algorithm to efficiently construct a metric satisfying this
requirement for large geographical areas. We have discussed how the geo fencing
technique can be elegantly expressed in the metric and incorporated in its
construction. Finally, we have performed an extensive evaluation of our
technique in Paris' wide metropolitan area, using two real-world datasets from
the Gowalla and Brightkite social networks. The results show that the adaptive
behavior of the elastic mechanism can offer better privacy in low-density
areas by adjusting the amount of applied noise.

\bibliographystyle{abbrv}
\bibliography{short}

\end{document}